\documentclass[12pt,a4paper]{article}
\usepackage{amsmath}
\usepackage{amsthm}
\usepackage{epsfig}
\usepackage{latexsym}
\usepackage{pdflscape}
\numberwithin{equation}{section}
\usepackage{graphicx}
\title{\Large \bf Day of the week effect in paper submission/acceptance/rejection to/in/by  peer review  journals  }
\author{ Marcel Ausloos$^{1,2,3}$, \\ Olgica Nedic$^{4}$, \\   Aleksandar Dekanski$^{5}$  }
 \date{$^1$  School of Management, University of Leicester, University Road, Leicester  LE1 7RH, UK;\\$e$-$mail$ $address$: ma683@le.ac.uk \\
 $^{2}$eHumanities
group\footnote{Associate Researcher}$\;$, \\Royal Netherlands
Academy of Arts and Sciences (NKVA), \\  Joan Muyskenweg 25, 1096 CJ
Amsterdam, The Netherlands \\ $e$-$mail$ $address$:marcel.ausloos@ehumanities.knaw.nl\\ 
$^3$GRAPES\footnote{Group
of Researchers for Applications of Physics in Economy and Sociology}$\;$,
  rue de la Belle Jardiniere 483, \\B-4031, Angleur, Belgium \\$e$-$mail$ $address$:
marcel.ausloos@ulg.ac.be
\\ \vskip0.5cm
$^4$ Institute for the Application of Nuclear Energy (INEP),
University of Belgrade, Banatska 31b, Belgrade-Zemun, Serbia
   \\$e$-$mail$ $address$: olgica@inep.co.rs \\  \vskip0.5cm
$^5$ Institute of Chemistry, Technology and Metallurgy,
Department of Electrochemistry, University of Belgrade, Njegoseva12, Belgrade, Serbia
 \\$e$-$mail$ $address$: dekanski@ihtm.bg.ac.rs  }
\begin{document}
 \maketitle

\clearpage
\begin{abstract}
This paper aims at providing an introduction to the behavior of authors submitting a paper to a scientific journal.
Dates of  electronic submission of papers to the Journal of the Serbian Chemical Society have been recorded from the 1st January 2013 till  the 31st December 2014, thus over 2 years.
  There is no Monday or Friday effect like in financial markets, but rather a Tuesday-Wednesday effect occurs:     papers  are more often submitted on Wednesday; however,  the relative number of going to be accepted papers is larger if these are submitted on Tuesday. On the other hand, weekend days (Saturday and Sunday) are not the best days to finalize and submit manuscripts. An interpretation based on the type of submitted work ("experimental chemistry") and on the influence of  (senior) coauthors is  presented. A thermodynamic connection is proposed within an entropy context.  A (new) entropic distance is defined in order to measure the "opaqueness"  = disorder) of the submission process.

\end{abstract}
   
  Keywords : scientific agent behavior;   submission day;  acceptance rate;  rejection rate; entropy; senior coauthors 
    \vskip0.5cm
    


 \section{Introduction}
Human behavior is controlled by many externalities. Much difficulty for selecting and analysing them resides in the complexity of establishing  experimental schemes, in a laboratory,  in view of mimicking the real world.  Comparison between laboratory outputs and  "common sense expectations" is often impaired by the lack of reliable (and unbiased) data  on  the latter.   For socio-physicists,  one  basic question about behavior is    geared toward finding whether there is some  unforced regularity in behavior,  e.g. weekly, monthly, seasonal or with longer time scales. For example, one finds quantitative considerations on human aspects of synchronized behavior or cyclic rhythms, like menstruation, heart beat  or birth rates \cite{bakbook}-\cite{HumRep}. 
  It is well known that there is a day-of-week effect in financial markets  \cite{French}-\cite{Cellini_AFE24.14.161}

 Here, we explore the behavior of  agents submitting papers for peer review and publication in a  scientific journal.  We find that, in the case at hand, more papers are submitted on Tuesday, but relatively (to the total number of submitted papers)  more papers are accepted for publications if submitted on Wednesday by the author(s).  On the other hand, the highest rejection rate  occurs for papers submitted on Saturday. Statistical tests are provided to ensure the validity of the findings.  
 These   were hardly predictable, whence it seems worth reporting them, even though they suggest  further questions and  investigations. While it would be interesting to compare our results with those obtained by studying other journals, such data is not easily available. 
We admit that it might be often possible  through IT infrastructure  to retrieve dates of submissions  of  accepted papers.  However, the case of rejected papers is rather hopeless:   such "trade secrets" are not  shared  with external researchers. We have been fortunate to get access to such data in the present case\footnote{ON and AD are Journal of the Serbian Chemical Society Sub-Editor
 and
Journal of the Serbian Chemical Society Manager, respectively}. 

 Our report is a case study.  In order to go beyond our observation, we are aware that   more data must be made available by editors and/or publishers. Nevertheless, the outlined findings and subsequent  hypotheses, short of a model we also agree on this,  on behaviors of    scientific authors, take into account their work environment. The results suggest some possibly universal feature for submitted manuscripts in so called "hard core science ".

 As an introduction to our aim,  recall that 
 "agents" can be considered  as  either rational or irrational, but are also  influenced by "external fields", in a socio-physics   sense  \cite{roehner2007driving}, 
 e.g. in stock market returns  \cite{DhesiWilmott} or book or record sales  \cite{PhA362sales RLMA}, respectively. 
It is logical to admit that  an agent  behavior is often  influenced by the action of others, but could also be intrinsic due to societal constraints or habit, - and memory experience as well \cite{PhA368.09.1849Lipowski}.  
For example, it  has been shown  that an investor behavior   is different,   
on Monday and Friday with respect to other days of the week, as noticed in many economist reports
 \cite{French}-\cite{Cellini_AFE24.14.161}, 
  about  stock exchange returns on markets in developed countries.  For completeness, let us acknowledge  that some debate goes on about such a behavior for stock markets pertaining  to "emerging  countries"   \cite{SaadiRahman,Kohersetal}. 
Here, we focus some attention on similar behavioral questions outside  the financial and economic sphere, but rather in a scientometrics framework. 

The best methodology should consider  strict investigation rules on whether  a "submission weekday behavior effect" exists and is confirmed, using specific data and  pertinent quantitative methods,  through questions as:
\begin{itemize} \item can one observe and analyze   (scientific) agent (behavior) effects?  \item   can one find out whether  there is any daily difference in behavior?
\item   can one  test whether the  findings are  reliable? 
\end{itemize}
 Thereafter some discussion is expected, together with suggestion for further investigations.

Due to the background of  the present researchers, they were able to obtain very  reliable data, see Sect. \ref{Data}, on  the submission of papers to a scientific journal.  Moreover,  the outcome, acceptance or rejection, was  provided. Whence we are able to measure whether some "day of the  week" effect  exists through statistical tests, - see  Sect. \ref{analysis}, not only on the most popular day for submission, but also on the day for which the outcome will be  the most positive (and the most negative) one.  Since the various probabilities of interest are empirically known, it is easy to connect the data description with the notion of entropy, thereby measuring and discussing the disorder of author submission of papers, as in Sect.  \ref{sec:entropy}. 
An interpretation of the findings is proposed in Sect. \ref{reasoning};  a short conclusion is made in Sect.  \ref {Conclusion}.

\section{Data}\label{Data}
\begin{figure}
\includegraphics[height=16.8cm,width=16.8cm]{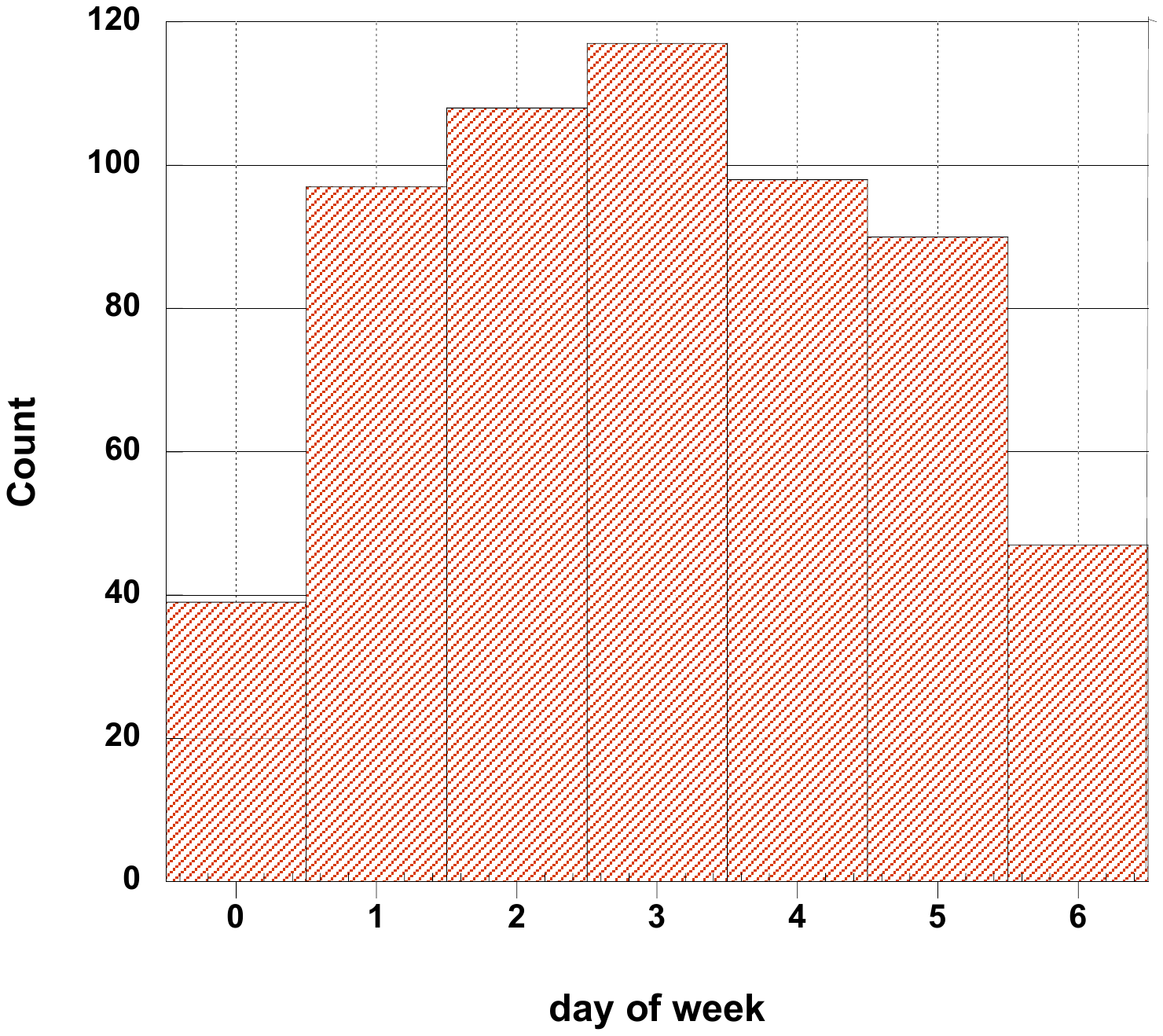} 
 \caption{ Number of papers submitted to JSCS according to the week day of submission in  2013 and 2014.} \label{Plot13Ndayofweek} 
\end{figure}

We have obtained data  about how many papers ($N= 596$, in fact) were   electronically submitted  to The  Journal of the Serbian Chemical Society   (JSCS)\footnote{http://shd.org.rs/JSCS/} for the years 2013 and 2014.  (The journal contains various sub-sections.)
 JSCS had  an impact factor= 0.912 in 2012. 
 A histogram of the day of the week submission is shown in Fig. \ref{Plot13Ndayofweek}.
 
Next, let us call  the numbers of papers later accepted ($N_a$) and those rejected ($N_r$);  the day-of-the week values are given  in Table \ref {JSCSNNaday}.    Among those 596,  $N_a= 262 $ were finally accepted for publication. At this time of writing,  2 are still under review, and 38 are not yet published. For completeness, let it be recorded that  $N_{nrr}=23$ were rejected because  the authors did not  reply to the reviewers remarks in due time, while  7 submissions were withdrawn ($N_w=7$). (Thus, $N_a+ N_r\neq N$). 
 \begin{table} \begin{center}
   \begin{tabular}{|c|c|c|c|c|c|c|c|c|c|}
      \hline           
       &   &	$N$	& 	$N_a$&	$N_r$& $N_a/N$& $N_r/N$&$N_{nrr} $&$N_w$ &$N_r/N_a$   \\
\hline 0&Sunday 	&	39	&	14	&	25	&	0.359	&	0.641	&	-&-&1.786\\
\hline 1&Monday	&	97	&	42	&	53	&	0.433	&	0.546	&	6&2&1.262\\
\hline2&Tuesday	&	108	&	60	&	46	&	0.556	&	0.426	&	5&1&0.767\\
\hline 3&Wednesday	&	117	&	51	&	64	& 0.436	&	0.547	&	4	&2&1.255\\
\hline 4&Thursday	&	98	&	43	&	54	&	0.439	&	0.551	&	3&-&1.256\\
\hline 5&Friday	&	90	&	38	&	50	&	0.422	&	0.556	&	 4&2&1.316\\
\hline 6&Saturday	&	47	&	14	&	33	&	0.298	&	0.702	&	2&-&2.357\\
\hline
\hline  &Total:	&	596	&	262	&	325	&	0.437	&	0.543	&	24&7&1.240 \\
\hline
\hline &$\chi^2$	&	62.354	&49.260	&22.868 &	-	&	-	&3.488&	5&-\\
\hline \end{tabular}   \end{center} 
\caption{Number $N$ of papers \underline{submitted on a given week day} to JSCS in 2013 and 2014, among which are the numbers of papers later accepted ($N_a$) or rejected ($N_r$). The relative percentages (per day) are also given. $N_{nrr}$ is the number of papers rejected because the authors did not reply to the reviewers' remarks in due time. $N_w$ are withdrawn submissions. The 95\% confidence of a null hypothesis (uniform distribution) for the 6 degrees of freedom $\chi^2$ = 12.592. 
 }    \label{JSCSNNaday}
\end{table}  %

 \begin{table} \begin{center}
      \begin{tabular}{|c|c|c|c|c|c|c|c|}
      \hline      
      &	$N$	& 	$N_a$&	$N_r$& $N_a/N$& $N_r/N$&$N_r/N_a$  \\
     \hline      \hline 
 Min.&39 	&	14	&	25 &	0.298	&	0.426	&	0.767	 
\\Max.&117	&	51	&	64 &	0.556	&	0.702	&	2.357	 
\\Sum&	599	&	262	&	325	&	2.92	&	3.93	&	9.998	 
\\Mean	&85.6	&	37.4	&	46.4	&	0.418	&	0.562	&	1.428	
\\Median	&98 	&	42	&	50 	&	0.429	&	0.551	&	1.262	
\\Std Dev.&29.7	&	17.5	&	13.3	&	0.082	&	0.0807	&	0.505
\\Skewn.&	-0.694	&	-0.383	&	-0.464	&	0.113	&	-0.0986	&	0.758	
\\Kurt.&	-1.12	&	-1.15	&	-0.864	&	-0.275	&	-0.218	&	-0.138	
 \\
\hline
 \end{tabular}   \end{center}  \label{JSCSNNadaystat}
\caption{Statistical characteristics of the distribution of the Number $N$ of papers \underline{submitted on a given week day} to JSCS in 2013 and 2014, and number of papers later accepted ($N_a$) or rejected ($N_r$). The characteristics of the percentage distributions are also given.}  
 \end{table} 
 
\begin{figure}
\includegraphics[height=16.8cm,width=16.8cm]{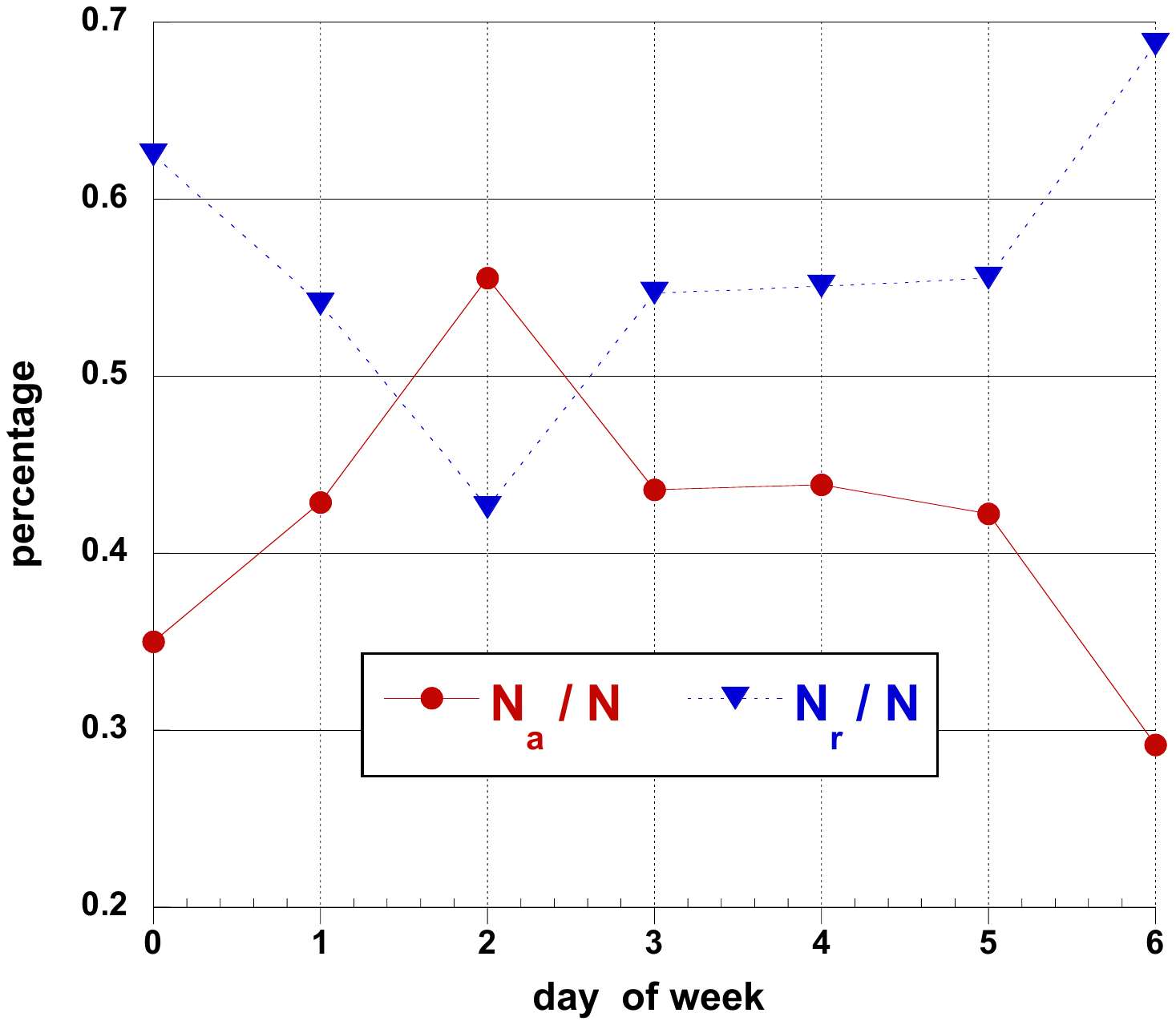} 
 \caption{ Percentages of accepted and rejected papers (with respect to the number of papers submitted on a given day)  according to the week day of submission  to JSCS   in  2013 and 2014.} \label{Plot3distanceplot} 
\end{figure}

  The relative number (expressed in percentages) of papers accepted or rejected after submission ($N_a/N$ and $N_r/N$)  on  a  specific day of the week is shown  in Table \ref{JSCSNNaday} and, for having a good
visual inspection, in Fig.  \ref{Plot3distanceplot}. It appears that, in contrast to the (more often occurring) submission day (Wednesday, day 3), - the next one being Tuesday, the papers are more often accepted when (or if) submitted on Tuesday (day  2). However, the largest number of rejected papers are those  apparently submitted on Wednesday (day  3). However, when expressed in relative terms (in percentages of the submitted papers on a given day), - unexpectedly, the greatest proportion of manuscripts gets rejected if submitted on  Saturday (day 6), - also on Sunday (day 0).   A Tuesday submission still  proportionally  remains the best day for a positive outcome.
  
 Consider the $N_r/N_a$ ratio  which actually emphasizes this rate of rejection. The ratio
decreases from 1.79 (Sunday), to 1.26 (Monday), goes to a minimum  0.77 (Tuesday), then increases 1.25, 1.26, 1.32 and 2.36 (Saturday); see Table \ref{JSCSNNaday}. It seems that  a Sunday-Saturday effect  can be extracted, if we want to judge the odds on rejection/acceptance probability, while a Tuesday-Wednesday effect is prominent for  a paper later acceptance.
  
 
 \section{Data analysis}\label{analysis}

First, observe that the summary statistics in  Table 2 show that the distribution of the daily submission, later accepted or not papers,   are  negatively skewed, - which has to be expected. The kurtosis of each distribution is also negative. However,  the  skewness of the distribution of percentages is not  always negative. 
 The results of the Doornik-Hansen test (based on the skewness and the kurtosis)   \cite{DoornikHansen} show that the empirical distributions of all the variables  are non-Gaussian. 

Next, recall also  that the  $\chi^2$ value at 0.95\% confidence is  18.5476 for 6  degrees of freedom for a uniform distribution.   Supposing that the distribution of submitted papers is week day independent, the calculated $\chi^2$  is  62.354, - see Table 1, thereby indicating that the distribution is far from uniform,  i.e.  there are significant differences about the day of the week. There is a  markedly  significant propensity to submit on Wednesday (day 3 of the week).

Supposing that the distribution of the accepted or rejected  papers is week day independent, the calculated $\chi^2$  are found to be equal to  49.260, and  22.868, respectively for  $N_a$, and $N_r$,  thereby indicating that the distribution is far from uniform,  i.e.  there are  significant differences about the day of the week positive (or negative) outcomes.

 
\section{Entropy connection}\label{sec:entropy}
 
The ratio between the number of papers submitted on a given day ($N_s^{(j)}/ N_T$) with $j$ indicating a day of the week $= 0,..., 6$,  and $N_T$ the total number of submitted papers  can be considered to be a  characteristic  (empirical) measure of the  daily probability $p_s^{(j)}$ of submission. Similarly   $N_a^{(j)}/N_a$ (and $N_r^{(j)}/N_r$) can be considered to be a  characteristic  (empirical) measure of the probability of  acceptance (or rejection)   for the paper submitted on a  given day, i.e.  $p_a^{(j)}$   and $p_r^{(j)}$  respectively.. 
  
Thereafter,  one can obtain something which looks like   a contribution to a Shannon information entropy  \cite{shannon48,shannon51}    for a given day  $j$,  e.g. for submission:
  \begin{equation} \label{Hj}
  H_s^{(j)}\equiv   -\;p_s^{(j)}   \; ln (p_s^{(j)} ),
  \end{equation}  
  leading to  $  H \equiv \;\sum_{j=0}^{6}  H^{(j)}$.  One obtains:  $H_s=1.8880$, 	$H_a=	1.8393$, and 	$H_r=	1.9082$, respectively. Such values have to be compared to the entropy of full disorder $H= ln (7)= 1.9459$, markedly higher than the others.
 Indeed, to estimate the validity of an empirical  distribution, it is  practical  to  compare each ($H^{(j)}$) measure 
  to their related maximum disorder number,  i.e.   $ln (N^{(j)})$. Thus,  we define the relative "distance"  to the maximum entropy (full disorder) as\footnote{The astute reader  observes that this definition is different from the "Kullback-Leibler divergence" measure \cite{Kullback}; the present one is  closer to a thermodynamic interpretation.} 
 \begin{equation} \label{d}d^{(j)}= 1-\frac {H^{(j)} }{ln(N^{(j)})} \end{equation}

   Eq.(\ref{d}) measures the information loss if the empirical probability ($p^{(j)}$)  is taken as an approximation  to that arising  from the  uniform distribution.  Then, $d^{(j)}$ contributes to a correct evaluation of the entity of the deviation of the set of data.   Therefore, $d^{(j)}$ can be interpreted as a proxy of the degree of opaqueness  (or of disorder) of the  "paper submission  market".  The $d^{(j)}$'s are respectively equal to  0.02975, 0.05478, and 0.01937 for the submitted, accepted and rejected cases.

   
\section{Interpretation}\label{reasoning}
Some reasoning on the findings can be proposed.  Much is due to the type of journal, type of papers, and type of authors! Recall that the journal is rather specialized, - in chemistry. It can be expected (and it is observed) that most papers are  reporting rather  experimental data, and are often co-authored. It should be of common knowledge that much work is produced every day of the week in an active laboratory, but much writing (after data analysis) occurs often during week-ends. Moreover, young researchers are not expected to bother their advisor during the week-end. Thus,  much discussion between them occur at the beginning of the week. The  senior author is prone to request some revision of the writing, inducing a delay in the submission process, thereby leading to a more frequent submission of papers on the middle of the week.  The higher frequency of acceptance of papers for the Tuesday submission is conjectured to occur because the paper is better written and does not need an extra day of revision before submission.

Concerning the effect of week-end submission, 
it can be conjectured that researchers who submit on Saturday or Sunday do so less willingly during the week  because they work under pressure (someone or themselves expect a manuscript to be sent by Monday, - an unwilling task for a weekend). Maybe these scientists are less eager to read very carefully once again their manuscript or  are less inspired to draw important conclusions or  to demonstrate  significant relations in the results. Moreover, since it is  remarkable that the probability of  a paper (later) acceptance ($N_a/N_r$)  is higher when  the paper is submitted on Tuesday, it seems that it is not worthwhile to wait an extra day (or more)  to polish a paper.

\begin{figure}
\includegraphics[height=16.8cm,width=16.8cm]{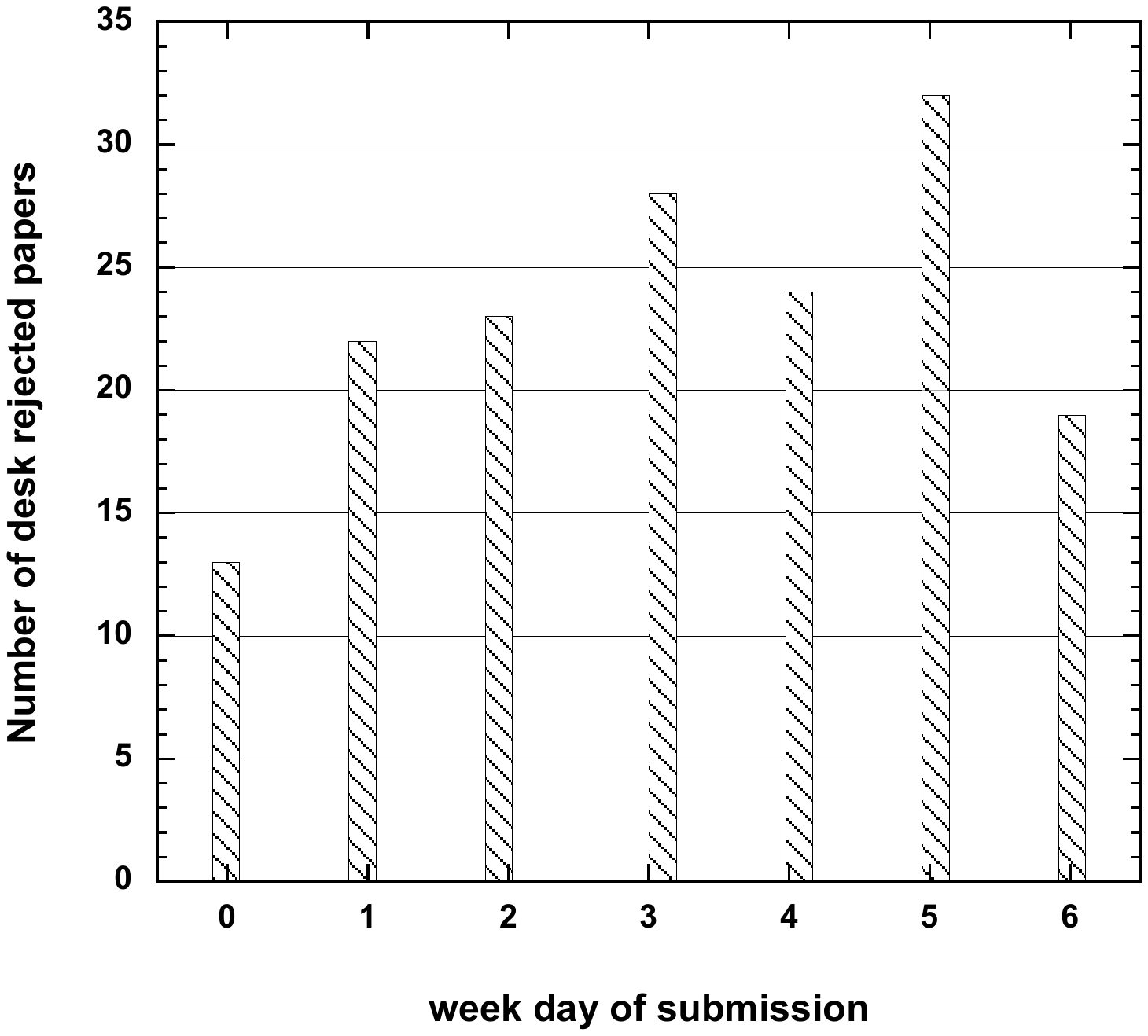} 
 \caption{ Distribution of desk rejected papers according to the week day of  (electronic) submission to JSCS   in  2013 and 2014.  The  $\chi^2$ test confirms the quasi uniform distribution. } \label{Plotdeskreject} 
\end{figure}

If necessary, we re-emphasize to the reader that the role of the  author is the concern: the role of the editor is irrelevant.  For example, one  might have imagined that the editor plays some role in the rejection of papers, yet electronically submitted by an author on a given day. On this remark, it can be observed that the number of desk rejected papers, i.e. $N_{dr}$ is equal to 161. Their  day of submission distribution is quasi uniform, as depicted on Fig. \ref{Plotdeskreject}; statistically, a $\chi^2$ test leads to  $\chi^2 \simeq 9.74$, quite below the $\chi^2  =12.59$, corresponding to the  95\% confidence of a null hypothesis (uniform distribution) for the 6 degrees of freedom.

\section{Conclusion}\label{Conclusion}

In summary, there is quite a number of studies on the "day of the week" effect on financial markets. To the best of our knowledge, this is the first time that one quantifies submission of scientific papers, thus the behavior of scientific agents in such a process, i.e. considering some author's brain work and scientific activity content. Moreover,  to take into  a specific outcome (later acceptance or rejection) does not seem to have been studied.
  
 It has also been shown that the analysis should take into account the relative size of daily submissions within a week. This normalization is relevant in order to observe whether the acceptance and rejection rates  (= probabilities) will differ depending on the day of submission. 
 
  According to our results, it seems that weekend days (Saturday and Sunday) are not the best time to finalize and submit manuscripts; it would be further intriguing to see how many of such  papers are desktop rejected compared to peer-review rejected. Of course, the fate of the manuscript highly depends on other participants in the peer-review and publishing process \cite{Hargens}, namely editors and reviewers. In order to reveal (possible) "day of the week" effect in the entire process of scientific publication, it would be of interest to investigate when reviewers are informed that they should review a paper, when they accept (or refuse) to review the submission, when they comment on the paper, and when editors finally accept or reject papers during the week, but such data for the JSCS are alas not available.

 Moreover, this type of investigation on editors and reviewers bears upon other sets than the one here investigated in the peer review process community. In fact, the role of editors and reviewers in slowing don or speeding up a review process (independently of the day of submission) can be studied. In   \cite{MFFNAeditwork}, we   discuss that the peer review process can be separated into distinguished  stages  for going
 through  the peer review process. We have introduced the notion of completion rate, -  a measure of the
probability that an invitation sent to a potential reviewer  results in a finished
review. Using empirical transition probabilities and probability distributions of
the duration of each stage, a directed weighted network can be created. Its  analysis
 allows   to obtain the theoretical probability distributions of review
time for different classes of reviewers. Through these simulations, we test the impact of some modifications of the editorial policy on the  efficiency  of the whole review process.  Practically, the results may act as a guide in
determining the optimal number of reviewers  \cite{MFFNAeditwork}.  However, it should be noticed that this data  analysis and subsequent modeling pertain to reviewers and editors behaviors, - not to authors as examined in the present paper. Nevertheless, day of week behavior of editors and reviewers might be interesting complementrary subjects to be considered  for a better transparency of the peer review process.
    
In conclusion, let  us to offer a few specific suggestions for further research lines on authors. 
It would be interesting to see: (i) whether, in other types of journal,  there is a general trend in authors' behaviour when choosing the submission day of the week, also  (ii) whether single author papers "behave" differently of co-authored papers,   (iii) if  "more theoretical" or "more experimental" papers "behave" differently, and (iv) whether the day of submission of a revised version is pertinent. Moreover, if the acceptance rate is due to a fine writing quality (and interesting results),  does the number of citations by others depend on the day of submission?  has a day of submission (and necessarily acceptance)   a real impact on science ?  


\vskip0.5cm

   \vskip0.5cm
 {\bf Acknowledgements.}
   This paper is part of scientific activities in COST Action  TD1306 New Frontiers of Peer Review (PEERE).

\vskip0.5cm

 \begin{figure}
\includegraphics[height=16.8cm,width=16.8cm]{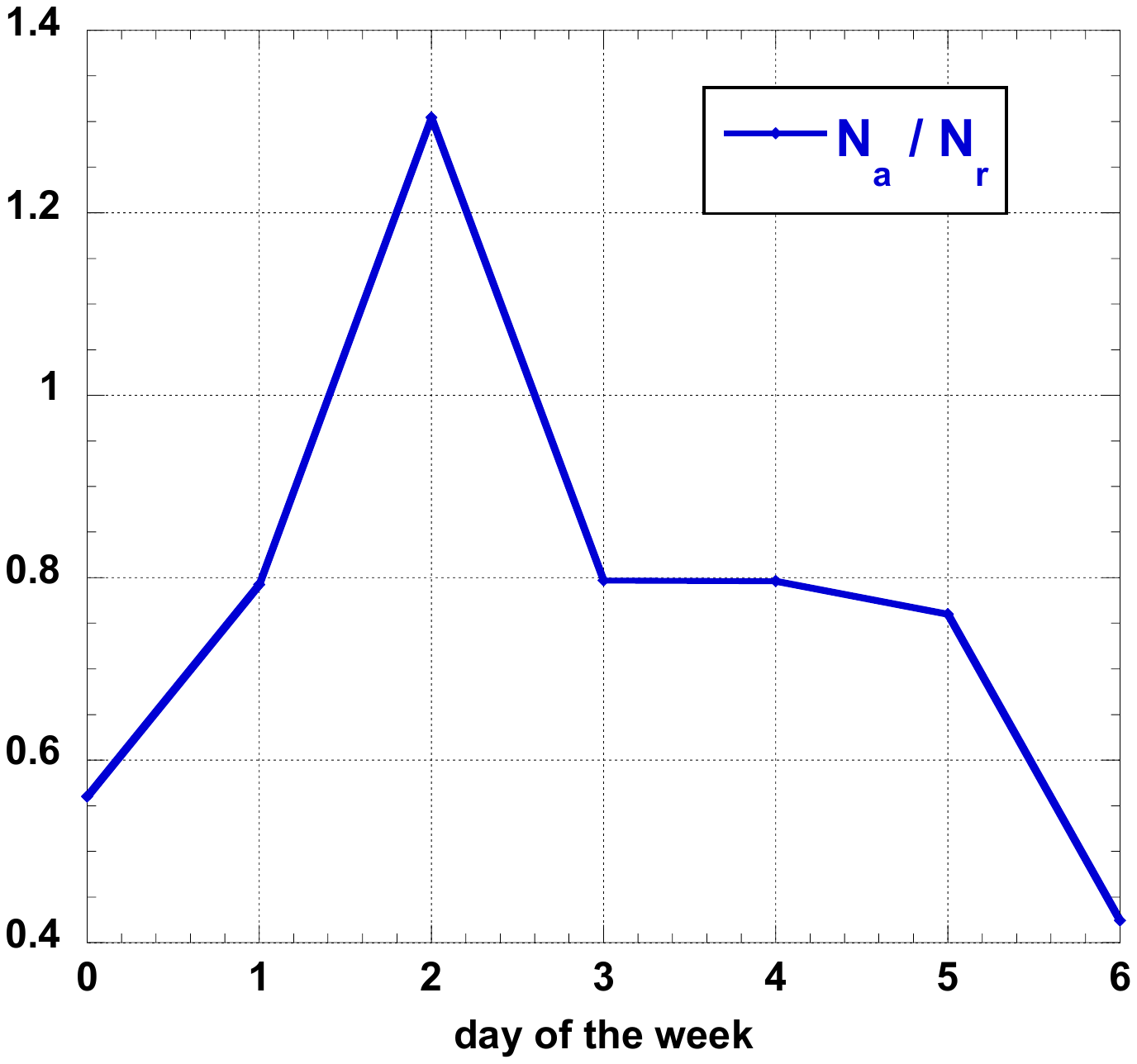} 
 \caption{ Ratio of  accepted to rejected  papers according to the week day of submission to JSCS   in  2013 and 2014. } \label{Plot5ratioNaNr} 
\end{figure}

\newpage


\end{document}